%% file: IoT.tex
\documentclass[lettersize,journal]{IEEEtran}
\input{math_commands}

\usepackage{amsmath, amsfonts, amssymb, amsthm, mathtools}
\usepackage{algorithmic}
\usepackage{algorithm}
\usepackage{hyperref} 
\usepackage{array}
\usepackage[caption=false,font=normalsize,labelfont=sf,textfont=sf]{subfig}
\usepackage{float}
\usepackage{textcomp}
\usepackage{stfloats}
\usepackage{url}
\usepackage{color}
\usepackage{parskip}
\usepackage{verbatim}
\usepackage{graphicx}
\usepackage{doi}
\usepackage{cite}

\begin{document}

\title{Try to Avoid Attacks: A Federated Data Sanitization Defense for Healthcare IoMT Systems}

\author{Chong Chen, Ying Gao,~\IEEEmembership{Member,~IEEE}, Leyu Shi, and  Siquan Huang
% <-this % stops a space
        
%%\thanks{Manuscript received August 30, 2022; revised ~,~.}

\thanks{
This work is supported in part by Guangdong Provincial Key
Laboratory of Artificial Intelligence in Medical Image Analysis and Application with No. 2022B1212010011, the Key-Area Research and Development Program of Guangdong Province under grants 2020B010166001, the Key Areas R\&D Program of Science and Technology Program of Guangzhou under grants 202103010005 and the National Natural Science Foundation of China with No.52102400.
%This paper was supported by the .\\
}
\thanks{Chong Chen, Ying Gao, Leyu Shi, Siquan Huang are with the School of Computer Science \& Engineering, South China University of Technology, Guangzhou 510006, China.(Email: {cschenchong,csshileyu,cssiquanhuang}@mail.scut.edu.cn; gaoying@scut.edu.cn)}
\thanks{Siquan Huang is also with the Guangdong Provincial Key Laboratory of Artificial Intelligence in Medical Image Analysis and Application, Guangzhou 510006, China.}
\thanks{\emph {Corresponding author}: Ying Gao.}
}
% <-this % stops a space}


%\markboth{IEEE Transactions on Computational Social Systems,~Vol.~?, No.~?, August~2022}%
%{Chen \MakeLowercase{\textit{et al.}}: Try to Avoid Attacks: A Federated Data Sanitization Defense for Healthcare IoMT Systems}


\maketitle

\begin{abstract}
Healthcare IoMT systems are becoming intelligent, miniaturized, and more integrated into daily life. As for the distributed devices in the IoMT, federated learning has become a topical area with cloud-based training procedures when meeting data security. However, the distribution of IoMT has the risk of protection from data poisoning attacks. Poisoned data can be fabricated by falsifying medical data, which urges a security defense to IoMT systems. Due to the lack of specific labels, the filtering of malicious data is a unique unsupervised scenario. One of the main challenges is finding robust data filtering methods for various poisoning attacks. This paper introduces a Federated Data Sanitization Defense, a novel approach to protect the system from data poisoning attacks. To solve this unsupervised problem, we first use federated learning to project all the data to the subspace domain, allowing unified feature mapping to be established since the data is stored locally. Then we adopt the federated clustering to re-group their features to clarify the poisoned data. The clustering is based on the consistent association of data and its semantics. After we get the clustering of the private data, we do the data sanitization with a simple yet efficient strategy. In the end, each device of distributed ImOT is enabled to filter malicious data according to federated data sanitization. Extensive experiments are conducted to evaluate the efficacy of the proposed defense method against data poisoning attacks. Further, we consider our approach in the different poisoning ratios and achieve a high Accuracy and a low attack success rate.

\end{abstract}


\section{Introduction}
%%ImOT应用前景（Wearable healthcare devices）
\IEEEPARstart{W}{earable} healthcare devices such as smartphones, wristbands, and smart glasses are highly efficient for medical monitoring, recording users’ health status by tracking activities. 
This widespread usage of innovative healthcare mobile and edge devices has formed the Internet of Medical Things (IoMT), which connects personnel and items related to medical facilities with the Internet to realize intelligent features of modern medical services{~\cite{FedHome,chen2020fedhealth, nguyen2022federated}}.

Healthcare IoMT Systems are designed to realize such as disease prediction, intervention, disease diagnosis, and treatment by collecting and analyzing patients' physiological parameters{\cite{TCSScai}. 
Wearable healthcare devices have become an essential role of IoMT Systems, empowering the IoMT to provide early warnings to several cognitive diseases, such as Parkinson’s and Heart Disease{~\cite{IoT_health, Gao}}.
%%ImOT智能应用服务场景中隐私安全问题
Since Healthcare IoMT Systems have increased the amount of proprietary user data in the medical field, the massive amount of data provides opportunities to develop machine learning models specific to various medical analytic domains and raises concerns about medical users' security and privacy{~\cite{IoT_FL}}. 
Thus the artificial intelligence models' training and application are impeded by the risk of privacy disclosure in Healthcare IoMT Systems.

%% 正常数据与带毒数据在模型推理阶段不同深度的特征之间的差别
\begin{figure}[!t]
\centering
\includegraphics[width=0.5 \textwidth ]{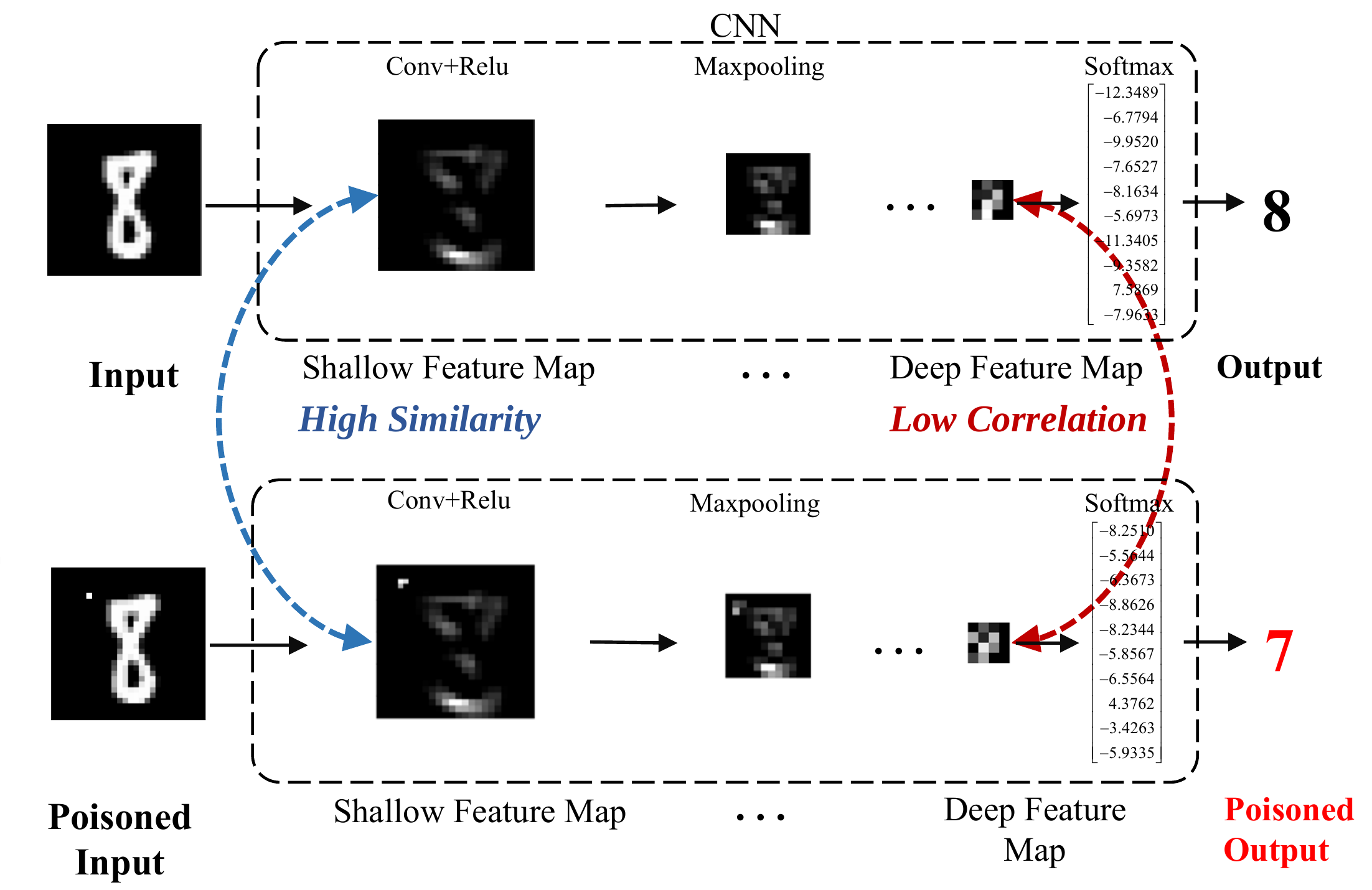}
\caption{Poisoned CNN responses to the normal data and poisoned data, the data "8" with its corresponding poisoned data "8" in MNIST database. There is a high similarity between input with poisoned input in the shallow feature maps and a low correlation to deep feature maps.}
\label{fig_1}
\end{figure}

%% 联邦学习简介
Federated learning has been proposed as a new data privacy-preserving decentralized learning paradigm{~\cite{mcmahan2017}}. 
By flowing the machine learning algorithm program to the parties that own the data, and then passing the training parameters back, this paradigm avails joint training of models through interactive model intermediate parameters to ensure data privacy, security, and legal compliance.
For Healthcare IoMT Systems, federated learning is a viable way to connect healthcare and bio-medicine data, allowing healthcare facilities to share experiences, not data, with guaranteed privacy{~\cite{Science2020federated}}.
Federated learning is used to establish a sharing learning architecture between mobile terminals and servers, so as to effectively use these data resources in the context of large-scale data and ensure the privacy of users{~\cite{yang2019federated,sun2020,Hu2020}}.
%% 联邦学习面临的攻击挑战
Among the scenarios, federated learning ensures and promotes data security and openness. 
However, the health-related IoMT services are so sensitive that federated learning models of any security breach would be catastrophic. 

Data poisoning attacks, also known as training data pollution, occur during the training stage of deep learning models{\cite{biggio2012poisoning}}.
When the attacker has the ability to obtain, modify or create new training data sets, he or she might poison the model which makes the training stage error by modifying a certain amount of poisoned data{~\cite{hayes2018contamination, fung2018mitigating, Sun2022 }}. Thus, the poisoned model learns the wrong correspondence designed by the attacker. Data poisoning attacks can target a part of the data with the same label or data with various labels{~\cite{cao2019understanding}}.

Data poisoning attack aims to change the probability distribution of the original training data so that the trained models, e.g. Convolutional Neural Network(CNN) models, make the error prediction{~\cite{ biggio2012poisoning}}. The most dangerous scenarios of data poisoning attacks are  kind of data backdoor attacks. The backdoor is activated by the pre-defined pixel pattern or semantic triggers in the inputs{~\cite{backdoor}}. 
By attacking "open the back door" to the target model, the abnormal behavior of the target model in some specific situations can benefit the attacker, and in general, the performance is the same as that of the normal model.
Fig.\ref {fig_1} briefly explains poisoned CNN responses to the normal data and poisoned data.
Data poisoning attack has been extensively studied in the centralized learning paradigm, causing concern in the field of healthcare security{~\cite{ mozaffari2014systematic, qayyum2020secure}}.
Since machine learning is vulnerable to data poisoning attacks, several methods have been proposed to intensify these attacks on federated learning{~\cite{cao2019understanding, hayes2018contamination, fung2018mitigating }}.
The more direct way to defend against the data poison attack is data sanitization, to filter out clean training data.
However, lacking supervision conditions makes data sanitization a challenging problem that requires an unsupervised learning approach to solve it.

%% 本文介绍
In this paper, we introduce a \textit{Federated Data Sanitization Defense} a novel approach for federated learning to avoid data poisoning attacks.
Federated Data Sanitization formulates clustering as the data sanitization strategy, which is solved the problem of a set of poison data to filter out clean data in federated learning{\cite{ghosh2020efficient, li2022secure}}.
Clustering is one of the unsupervised learning tasks and has been extensively studied in both centralized and distributed settings~\cite{dhillon2002data,tasoulis2004unsupervised,smith2017federated,ghosh2019robust,ghosh2020,sattler2020clustered}.
Constructing clustering in federated learning as a defense against data poisoning attacks can be regarded as a class of bilevel optimization problems with poison and clean data.
The data sanitization strategy relies on the fact that poisoned data are often outside the expected input distribution. Suppose a piece of poisoned data has a similar shallow feature to the normal one. In that case, the labels which stand for the deep feature must be different and vice versa, e.g.(For images, the shallow feature can be the pixel information and the deep feature be the semantics). Therefore, poisoned data can be treated as outliers, and data sanitization (i.e. attack detection and removal) can be used.
The illustration of our method is shown in fig.\ref {step}. 
Our main contributions are summarized as follows:
\begin{itemize}
\item{A Federated Data Sanitization Defense is proposed, our method protects training data and is dependable in federated learning, avoiding data poisoning attacks. }
\item{Our method achieves federated clustering to identify the poisoned data, the clustering is based on the consistent association of data and its semantics.}
\item{As an unsupervised method, our work does not rely on prior knowledge of the specific attack, e.g.,label-flipping attacks, backdoor attacks, or hybrid attacks, yet achieves a good performance.}
\item{We further evaluate our method in the different ratios of the attacker, our method achieves high accuracy and rather a low attack success rate.}
\end{itemize}

\begin{figure}[!t]
\centering
\includegraphics[width=0.5 \textwidth]{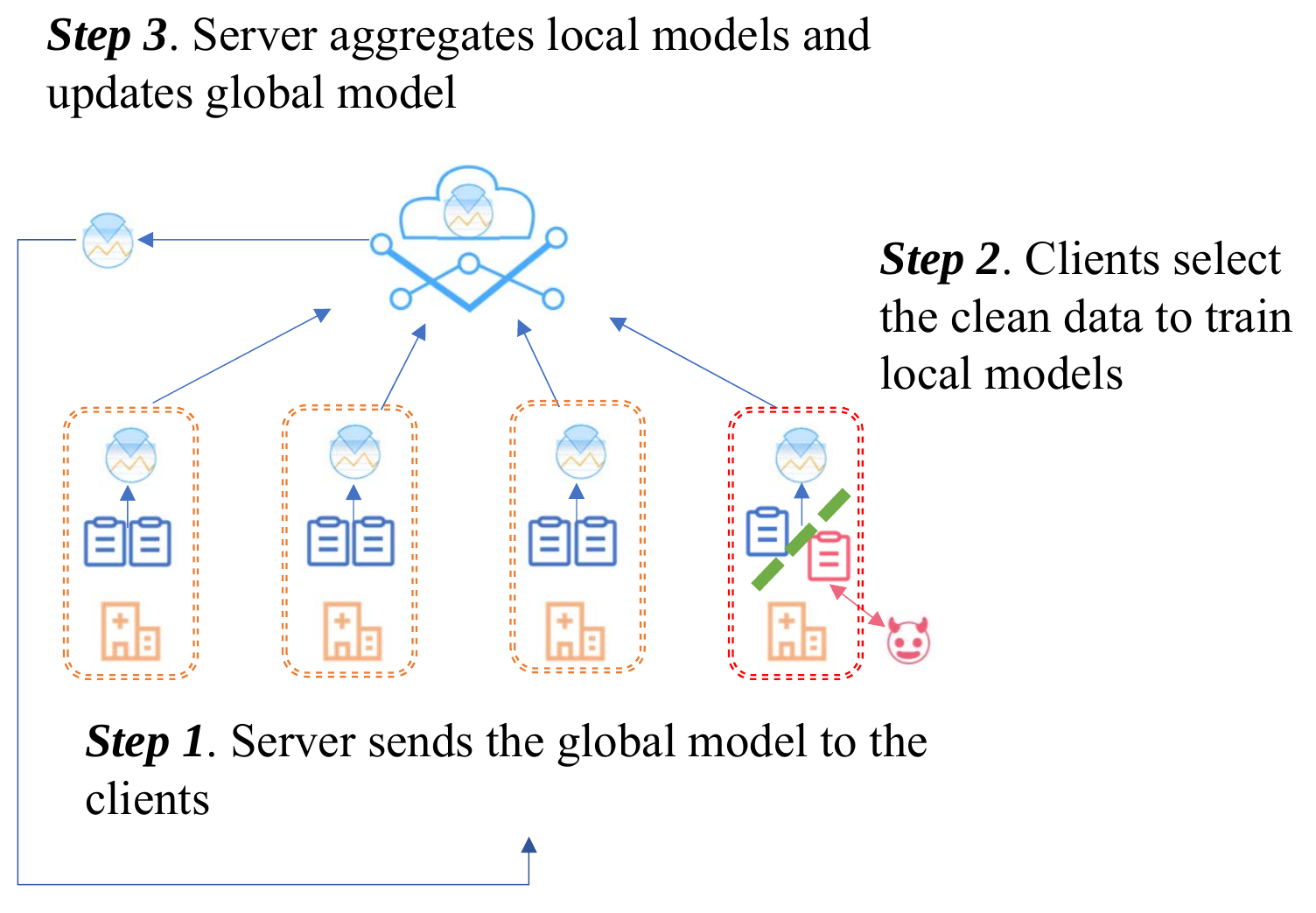}
\caption{A Federated Data Sanitization Defense for Healthcare IoMT Systems, to defend against the data poisoning attacks, here are the three main steps of Federated Data Sanitization.}
\label{step}
\end{figure}

\begin{figure*}[t]
\centering
\includegraphics[width=\textwidth]{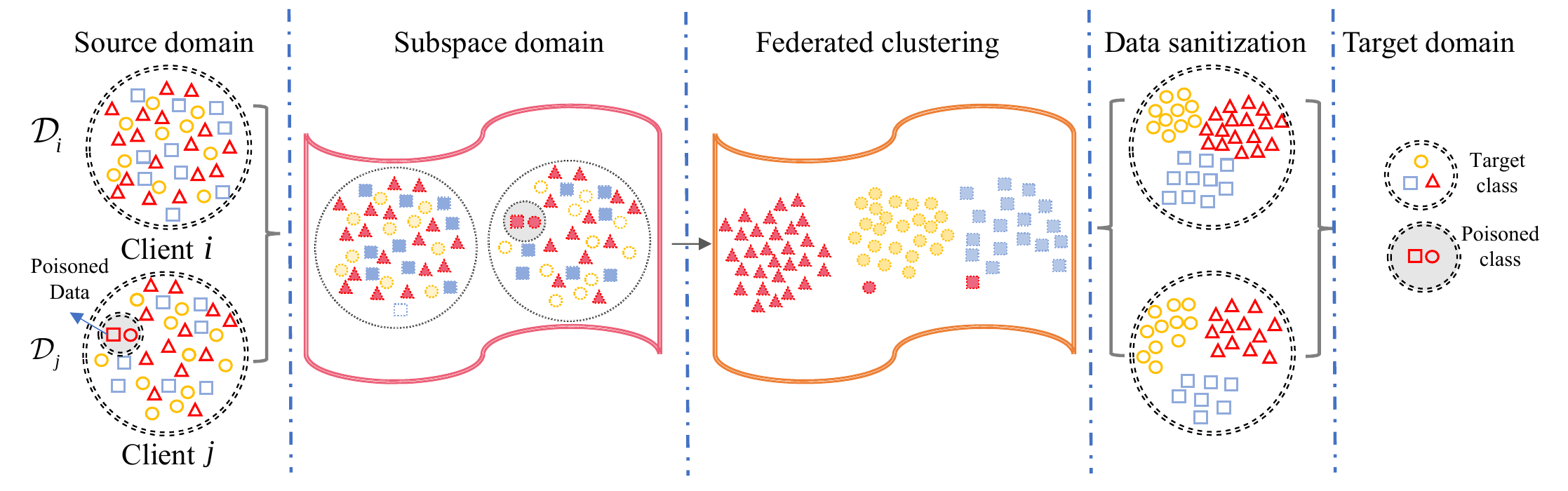}
\caption{Architecture of our proposed Federated Data Sanitization Defense. The purpose of our method is to remove malicious data and ensure the training data is safe for the model in federated learning. }
\label{fig_2}
\end{figure*}

\section{related work}
\subsection{Federated Learning in ImOT}
The lack of high-quality labeled data has become a stumbling block restricting the development of healthcare services. Based on an innovative operating concept, federated learning has enabled many attractive advancements in smart healthcare{\cite{IoT_health}}.
FedHome proposed a hierarchical joint edge learning framework, where the model aggregation part is done by directly affiliated hospitals, improving the In-home health monitoring services{\cite{FedHome}}.
FedHealth utilized transfer learning methods to achieve personalized model learning{\cite{chen2020fedhealth}}.
Federated learning also is a viable approach to linking healthcare facility Electroencephalogram(EEG) signals, allowing healthcare facilities to share experiences, not accessing the raw data, with guaranteed privacy{\cite{TCSScai}}. 
Federated learning allows service providers to train models without accessing data, but potential adversaries on distributed clients can use data poisoning to maliciously damage federated learning models.

\subsection{Federated Clustering}

Clusters are divided according to the similarity of clients, and each group is optimized by federated learning, which can enormously improve the performance and efficiency{~\cite{tasoulis2004unsupervised, kumarkannan01, smith2017federated,ghosh2019robust}}.
Since clustering is widely-used for unsupervised learning tasks, a variety of clustering methods have been proposed for non-independent and identically distributed(non-IID) data  in federated learning.
Based on the assumption that there is a reasonable partition so that the subset obtained after clustering satisfies the classical federated learning assumption(transforming the non-IID problem to an IID problem), Clustered Federated Learning proposes a framework for federated client group learning, in which the parameter server performs dynamic division according to participant gradients or update information{\cite{sattler2020clustered}.
The iterative federated clustering algorithm is implemented by the client's cluster membership, which learns and optimizes the clustering model for clients with similar data distributions~\cite{ghosh2020}.
Further, the $\rk$-FED method analyzes a one-shot communication scheme for federated clustering, using Lloyd’s method for $\rk$-means clustering, and proves heterogeneity can have distinct benefits for a subset of problems
\cite{dennis2021heterogeneity}.

\subsection{Defenses Against Data Poisoning Attacks}

Data poisoning attacks aim to compromise feature selection which was crucial to learning methods{\cite{xiao2015feature}}. 
Due to the distributed training data in federated learning are not released to the trusted authority and checked, one of the major threats to federated learning is that training data would be mixed with maliciously poisoned data—whether inherent or purposefully inserted by a malicious party, causing the trained model to misclassify{~\cite{ biggio2012poisoning, li2016data, alfeld2016data, wang2021robust}}.
In order to avoid attacks, one effective way is to detect malicious samples from the perspective of protecting the model's training stage. 
BayesAdapter adapts the pre-trained deterministic Bayesian
neural networks via cost-effective Bayesian fine-tuning
to detect the adversary samples{\cite{deng2020bayesadapter}}, and LiBRe appropriate for adversarial detection is proposed to boost the effectiveness and efficiency of BayesAdapter by building the few-layer deep ensemble variational and adopting the pre-training and fine-tuning workflow{\cite{deng2021libre}.

As the defense against poisoning attacks, another way is to improve the robustness of federated learning by selecting models.
The server eliminates a small part of models with large differences between a set of client models by choosing an appropriate metric, e.g.(Krum{\cite{blanchard2017machine}}, TrimmedMean{\cite{yin2018byzantine}}).
The defense limits the impact on poisoned data by selecting a port of the client models for model aggregation and eliminating models that are damaged by malicious data attacks{\cite{ blanchard2017machine, yin2018byzantine, guo2022adfl}}. 
The drawback of selecting models is that they can be computationally expensive for large models and the number of clients compared to standard aggregation rules.
And this kind of method reduces the aggregation model's accuracy to some extent{\cite{zhang2021robustfl}}.

\section{Federated Data Sanitization Defense}

In this section, we first present the threat model of data poisoning attacks and the defense problem formulation.
For a better understanding of the problem formulation, we introduce the problem from the perspectives of both the adversary and the defender. 
Then we detailed the proposed \textit{Federated Data Sanitization Defense} against the attacks.

\subsection{Problem Formulation}
Given the training dataset $\gD=\bigcup_{i=1}^N\{ (\bm{x}_i,y_i )\}, i\in\{1,\cdots, N\}$, each clients owns its private dataset $\gD_i$, in which $\bm{x}_i\in[0,1]^d$ is an image data and $y_i\in\{1,\cdots,\gK\}$ is the ground truth laybel of the image $\bm{x}_i$.
As for data poisoning attacks, especially backdoor attacks, the adversary can embed backdoor triggers during the training stage by modifying a certain proportion of training samples as the poisoning dataset in the federated learning.
The generic form of a poisoned image $\bm{x}^{'}$ with a hidden trigger as eq.\ref{poisoned_data} in which $\gP$ is  the transfer function to apply the trigger on $\bm{x}$,  $\bm{m}\in\{0,1\}^d$ is a binary mask of trigger positions, $\bm{p}\in\{0,1\}^d$ is the backdoor pattern\cite{dong2021black}.  

\begin{equation}
\begin{aligned}
\bm{x}^{'} &= \gP(\bm{x},\bm{m},\bm{p})\\
   &= (1-\bm{m})\cdot{\bm{x}}+\bm{m}\cdot{\bm{p}} 
\end{aligned}
\label{poisoned_data}
\end{equation}

Assume that the adversary has prior knowledge of the $\gD_i$, and creates malicious poisoned data injected into it. Let $\gD_i^{'}\subset\gD$ be the poisoned dataset $\gD_{i,p}^{'}=\{{(\bm{x}_i,y_i)} \mid\gP(\bm{x}_i,\bm{m},\bm{p}),y_i^{'}=y^t,(\bm{x}_i,y_i)\in\gD_i^{'}\}$, in which $y^t$ is the backdoor-target class.
Thus, the poisoned training dataset for federated learning is completed as $(\gD\setminus\gD^{'})\cup \gD_{i,p}^{'}$.
As for another kind of data poisoning attack, the label flipping attack only inverts its label to another chosen label.

Although the defender has no knowledge of the poisoned data to defend against these attacks. For a realistic scenario of 
data poisoning attacks in federated learning, the adversary is not going to transform all the data into poisoned data since he or she can not have control over the data of all clients in federated learning\cite{biggio2012poisoning,cao2019understanding, hayes2018contamination, fung2018mitigating}. In this consideration, the defender gets the knowledge that the adversary will only choose a small part of data of one class to create poisoned data, so the data on the clients can be re-clustered using data sanitization to filter out the poisoned data. 
Here we use $\{U_i^{(1)},U_i^{(2)},\cdots,U_i^{(k)}\}, k = card \{y_i \mid (\bm{x}_i,y_i )\in {\gD_i}\}$ as the clusters of the client $\gD_i$, the details of the federated clustering is in Sec.\ref{federated_clustering}. 
After the federated clustering forms a disjoint partition $\{U_i^{(1)},U_i^{(2)},\cdots,U_i^{(k)}\} $for every client’s training data ${\gD_i}$. Each client can filter out its poisoned data locally, since the shallow feature maps ${\mA_i}$ of the data ${\gD_i}$ have a ground-truth label.

\subsection{Federated Clustering}
\label{federated_clustering}

Federated Clustering provides a simplicity approach to cluster the no generative assumptions on the distributed data{~\cite{smith2017federated,ghosh2019robust}}.
Federated learning enables distributed data to map data features into a subspace domain{\cite{yang2019federated}. 
Because the poisoned data and normal data have similar shallow features and low correlation of data semantics. When clustering these shallow features by k-means or other clustering methods, the poisoned data will form the same cluster as its origin class data. Thus it will be eliminated through the data semantics: the ground-truth label{\cite{kumarkannan01}}.
Federated Clustering consists of two major stages, the local $k$-means clustering and federated aggregation{\cite{dennis2021heterogeneity}}.

\begin{algorithm}[tb]
  \caption{Local $k$-means clustering}
\label{alg_kmeans}
\begin{algorithmic}[1]
  \STATE {\bfseries Input:} The clients' data $\gD_i$ indexed by $i$, the matrix of shallow features in data $\{\mA_i\}$, class number of each client's data $k_i$;
  \STATE Project $\{\mA_i\}$ and run the standard approximation algorithm to estimate $k_{i}$ centers ($\nu_1,
  \nu_2, \dots, \nu_{k_{i}}$);
  \STATE Partition the $\{\mA_i\}$ into $k$ disjoint groups \\
  $\{\mU_i^{(1)}\mU_i^{(2)},\cdots,\mU_i^{(k)}\}$;
  \STATE Calculate $\tilde\Delta_{c},c\in{k_i}$ by eq.{\ref{subset}}, and verify satisfying eq.{\ref{separation}}
  \STATE {\bfseries Return:} Cluster assignments $(\mU_i^{(1)}\mU_i^{(2)},\cdots,\mU_i^{(k)})$ and their means $\Theta_{i} = (\theta^{(1)}_i, \dots, \theta_i^{(k)})$.
\end{algorithmic}
\end{algorithm}

\textbf{Local $k$-means clustering.}
Given a set of matrix $\{\mA_i\},\mA\in\mathbb{R}^{n\times{d}}$ is the shallow feature map data of the $\{\gD_i\}$. Let $\Vert{\mA}\Vert$ denote the norm of a matrix $\mA$, 
$\{c(\mA)\mid \mA_i \in {U_i^{(c)}}\}$
is the cluster index for $\mA_i$.

Assume the defenders have the prior knowledge about $k$ as the class number of $\{\gD_i\}$. For local $k$-means clustering of the $\{\mA_i\}$, we partition the $\{\mA_i\}$ into $k$ disjoint groups $\gU_i=\{\mU_i^{(1)}\mU_i^{(2)},\cdots,\mU_i^{(k)}\}$, note that some of $\mU_i^{(c)}$ could be $\emptyset$. We use $\mu(\gS) = \frac{1}{\Vert{\gS}\Vert}{\sum_{i\in{\gS}}\mA_i}$ indicate the mean of data indexed by $\gS$, and $\mu_c = \mu({\mU_i^{(c)}})$ for short, the cost of minimize the $k$-means as the eq.{\ref{cost}}.

\begin{equation}
     \phi(\gU_i) = \sum_{j=1}^k \sum_{i \in \mU_j}\Vert{A_i-\mu(\mU_i^j)}\Vert_2^2 \,.
    \label{cost}
\end{equation}

For a subset $\mU_i^c\in\mA_i$, 
let $d_{c}(x)$ donate the distance of the data x to the set $\mU_i^c$. Define $d_{c}(x)=\min_{b\in{\mU_i^c}}\Vert a-b \Vert_2$, and we have the function of $\mA_i$ and the $\mU_i^c$  as the eq.{\ref{subset}} .

\begin{equation}
    \tilde\Delta_{c}=\sqrt{k}\frac{\Vert{\mA_i}-\bigcup{\mU_i^c}\Vert}{\sqrt{\vert \mU_i^c \vert}} \, .
    \label{subset}
\end{equation}

It has proved that the two clusters $\mU_i^c$ and $\mU_i^s$ are well separated if for large enough constant $m$ satisfies eq.{\ref{separation}}{\cite{dennis2021heterogeneity}}.

\begin{equation}
   \Vert{\mu_r - \mu_s} \Vert_2 \ge m(\tilde\Delta_r + \tilde\Delta_s) \, .
  \label{separation}
\end{equation}

\textbf{Federated aggregation.}
After clients complete the local clustering stage and compute
the cluster means. The server aggregates and accumulates these results to return the aggregated clustering means.
Clients $i\in[N]$ use algorithm {\ref{alg_kmeans}} to achieve the local clustering with the its data $\gD_i$. After that, clients send the cluster server $\Theta_{i} = (\theta^{(1)}_i, \dots, \theta_i^{(k)})$ and their clusters $(\mU_i^{(1)}\mU_i^{(2)},\cdots,\mU_i^{(k)})$.

Let $\{\mu_c\},c\in{\gK}$ be the server aggregated results,
and define the $\mU_c^{'}= \{c \text{ : } \mA_i^c \in \mU_i^s \text{ and } \theta_i^s\in \mu_c, s\in[{k_i}] \}$. Define $\bar{\theta} \leftarrow \argmax{ d_{c}(x)}$, 
and $\mU_i^c \leftarrow \mU_i^c \cup \{\bar{\theta}\}$.
Each round of partition $\theta^{(c)}_i$, $(i \in [N], c \in [k])$ into $k$ clusters, $(\mu_1, \mu_2, \dots,
  \mu_k)$ with $\mU_i^c$ as initial centers.
The cluster server aggregates and forms a disjoint partition for federated learning.

\begin{algorithm}[tb]
 \caption{Federated aggregation}
 \label{algo:fed}
\begin{algorithmic}[1]
  \STATE On each client $i\in [N]$, run algorithm \ref{alg_kmeans} with local data $A_i$ and
  $k_i$ and obtain device cluster centers $\Theta_i = (\theta_i^{(1)};
  \dots, \theta_i^{(k)})$ at the central node.
  \STATE For $i \in [N]$ and let ${\mU_i^c} \leftarrow \Theta_{i}$;
   \STATE Let $\bar{\theta} \leftarrow \argmax{ d_{c}(x)}$;
   \STATE $\mU_i^c \leftarrow \mU_i^c \cup \{\bar{\theta}\}$;
  \STATE Run one round to cluster $\theta^{(c)}_i$;
  $i \in [N], c \in [k]$ into $k$ clusters, $(\mu_1, \mu_2, \dots,
  \mu_k)$. Use points in $\mU_i^c$ as initial centers.
  \STATE{\bfseries Return:} the clustering $\{U_i^{(1)},U_i^{(2)},\cdots,U_i^{(k)}\} $ the clients cluster server and the corresponding federated aggregation.
\end{algorithmic}
\end{algorithm}

\begin{algorithm}[tb]
 \caption{Federated Data Sanitization Defense}
 \label{algo:defense}
\begin{algorithmic}[1]
  \STATE On each client $i\in [N]$, projects data $\gD_i$ to subspace and gets shallow feature maps ${\mA_i}$.
  \STATE
  Run algorithm \ref{alg_kmeans} with local data $A_i$ and
  $k_i$ and obtain device cluster centers $\Theta_i = (\theta_i^{(1)};
  \dots, \theta_i^{(k)})$ at the central node.
  \STATE Run algorithm \ref{algo:fed} the clustering $(\mu_1, \mu_2, \dots, \mu_k)$  the clients cluster server and the corresponding federated aggregation $\{U_i^{(1)},U_i^{(2)},\cdots,U_i^{(k)}\}$.
  \STATE Set  $U_i^{(k)}$ with the label $y^{U_c} = {{y_i}\mid(\bm{x}_i,y_i)\in Mo(U_i^{(c)})}$, do data Sanitization in the cluster.
  \STATE{\bfseries Return:} Clients with the safe data $(\gD^{'}\setminus \gD_{i,p}^{'})$, and the model trained by FedAvg.
\end{algorithmic}
\end{algorithm}

\subsection{Data Sanitization}
After all clients
has its $\{U_i^{(1)},U_i^{(2)},\cdots,U_i^{(k)}\}, \gD_i = \bigcup_{c=1}^k\{ U_i^{(c)} \}, c\in\{1,\cdots, k\}$, then we can set every $U_i^{(k)}$ with the label $y^{U_c} = {{y_i}\mid(\bm{x}_i,y_i)\in Mo(U_i^{(c)})},c\in\{1,\cdots, k\}$ which is the modal element in the cluster by the data labels. According to the cluster label by a majority vote, the data in the same cluster can be compared to determine whether it is poisoned data. The majority of data in the same class tend to be the same cluster, the defender can use the $y^{U_c}$ as the criterion
to select the data in the cluster and regard other data as poisoned data.

In the end, the server randomly selects a subset of local agents $S_t$. The number of models selected at each round is $K =|S_t|=\beta*N$ where $\beta$ is the percentage of participating clients and $N$ is the number of total clients. Each selected local agent $i \in S_t$ receives the current global model $G^t$ and trains it with his local data using stochastic gradient descent (SGD), obtaining a locally updated model $L_i^{t+1}$. The agent sends its model update $\delta_i^{t+1} = L_i^{t+1} - G^t$ back to the server. 
The server then aggregates the agents' model updates and adjusts the global model with the learning rate $\eta$ based on FedAvg{\cite{mcmahan2017communication}}.
\begin{equation}
    \label{eq:fl_update}
    G^{t+1} = G^t + \frac{\eta}{N} \sum_{i \in S_t} (L_i^{t+1} - G^t)\, . 
\end{equation}

\section{Experiment}
As the defense against data poisoning attacks, we choose federated clustering to do data sanitization. Since we have the prior knowledge that the MNIST is a $10$ classes dataset and the ChestMNIST is the $8$, so we choose the cluster number as $k_i=10$ and  $k_j=8$ relatively. After the clients in federated learning have completed the clustering of their private data $\gD_i$ into $10$ clusters $\{U_i^{(1)},U_i^{(2)},\cdots,U_i^{(k)}\}$, then every cluster choose the mode of the MNIST class. The chosen data is going to use as the clear data participating in federated learning.
Our method completes with the AUROR, FoolsGold, and Krum against the data poisoning attack,
in order to more comprehensively evaluate the Generality of our method to different poisoning attacks, we further mix label flipping attacks and backdoor attacks as the new hybrid attack.

\subsection{Datasets and Experimental Setting}
\textbf{Datasets.}
To verify if our method could effectively and efficiently equip the federated learning with principled data sanitization ability in various data poisoning attacks scenarios,
our method is evaluated in the widely used MNIST{\footnote{
\href{MNIST}{http://yann.lecun.com/exdb/mnist/}}}, and MedMNIST{\footnote{
\href{MEDMNIST}{https://medmnist.com/}}} datasets.
The MNIST database of handwritten digits has a training set of 60,000 examples and a test set of $10,000$ examples. Each image contains $28*28$ pixels with a single channel. 
The digits have been size-normalized and centered in a fixed-size image.
Like the MNIST dataset, the MedMNIST dataset performs classification tasks on lightweight $28*28$ images covering major medical image modalities and diverse data sizes, which contains a total of 10 pre-processed open medical image datasets (with data from multiple different data sources and pre-processed){\cite{medmnistv2}}, we use the ChestMNIST, which is the part of MedMNIST, to test our method.

\textbf{Experimental setting.}
We evaluated our method with varying poisoned methods and different attack strategies.
We choose a simple convolution neural network for efficiency.
As for MNIST, the network is constructed by two convolution layers, a max-pooling layer, and a fully-connected layer.
And for ChestMNIST, we use the ResNet18 as the base network{\cite{resnet}.
Our models run for federated learning architecture with $N = 10$ clients. 

Generally, the adversary's goal is to discriminately poison global models for a high attack success rate in order to victimize selected classes.
We also consider adversaries that can evade the entire training cycle and induce insidious behavior by crafting different poisoning attacks.
So we list our defense target:

\begin{itemize}
  \item{Label Flipping Attacks: In this form of attack, we choose a specific class to poison by inverting its own label to another cheating label.}
  \item{Backdoor Attacks: We use the trigger, e.g.(shape, size, and position), using the backdoor function to generate adversarial data. }
  \item{Hybird Attacks: As the combination of the two above attacks, we mix up the two poisoned data to attack the model. }
  \end{itemize}

In each round of training, the server randomly selects $N_k = 8$ clients. 
In order to highlight the effectiveness and Generality, we make a detailed comparison with existing methods by comparing the attack success rate and model accuracy rate of the model against malicious data. 
We measure efficiency by evaluating the execution of each step of the model.

\textbf{Data poisoned attacks.}
According to the target of the attack, 
Data poisoning attacks can be divided into random attacks and targeted attacks.
Random attacks: Reduce the accuracy of federated learning models, e.g.(label-flipping attack).
Targeted attack: Induce the federated learning model to output the target label specified 
by the adversary, e.g(backdoor attacks).

\textbf{Baselines.}
Since our method is a data sanitization way to clear the poisoned data, it is a thorough solution to the problem and prevents the federated learning models
from the poisoned data.
And there are some methods tries to make the model more generality against attacks.
One line of research aims to distinguish and exclude malicious model updates from the server aggregation. 
AUROR{~\cite{auror}} clusters the local model updates into two groups based on the identified informative features and only uses the majority cluster to update the global model.  
Another line of research focuses on designing a robust aggregation rule for FL systems.
Krum{~\cite{blanchard2017machine}} has a Byzantine-resilient aggregation rule that only selects the local model update with the smallest pairwise distance from the closest local updates. 
FoolsGold{~\cite{fung2018mitigating}} assigns different learning rates to the selected agents based on the pairwise cosine similarity between the local model updates to mitigate the influence of backdoor adversaries. 

\textbf{Metric.}
We estimate the accuracy of the attacking success rate of the defense model. Higher accuracy means the defense strategies do not influence 
the main task(model prediction of the normal data), and lower attacking success rate means the defense strategies
have a more positive capability against attacks.

We have evaluated our methods in the MNIST and the ChestMNIST against the attacks.
In the MNIST, we perform our model in the simple structure CNN for efficiency. 
And as for ChestMNIST, we choose the ResNet18 for better classification performance.
The experiments are conducted by Pytorch $1.6$ on a server computer that has ${4\times}$ Nvidia RTX 2080Ti, allowing all the clients to virtualize into it to train the federated model.

\begin{table*}[t]
\caption{Defense against three types of data poisoning attacks on the MNIST dataset, \\here we choose three layer CNN as the base network.}
\label{MNIST}
\centering
\begin{tabular}{cccc}
\hline
Methods (Accuracy/Attack success rate) & {label flipping }                     & {backdoor attack }                    & {hybrid attack }                      \\ \hline
AUROR{\cite{auror} }    & 88.46\%/35.30\%                       & 80.13\%/22.48\%                       & 82.53\%/41.36\%                       \\
FoolsGold{\cite{fung2018mitigating} }& 90.43\%/14.39\%                       & 82.61\%/18.54\%                       & 84.22\%/14.76\%                       \\
Krum{\cite{mcmahan2017communication} }     & 90.67\%/11.42\%                       & 81.54\%/{\textcolor[rgb]{1,0,0}{3.47\%}}                        & 84.46\%/9.67\%                        \\
Our method      & {\textcolor[rgb]{1,0,0} {92.43\%/2.32\%}} & {\textcolor[rgb]{1,0,0} {86.23\%}}/4.33\% & {\textcolor[rgb]{1,0,0} {87.35\%/5.24\%}} \\ \hline
\end{tabular}
\end{table*}

\begin{table*}[t]
\caption{Defense against three types of data poisoning attacks on the ChestMNIST dataset,\\
here we choose ResNet18 as the base network.}
\label{MedMNIST}
\centering
\begin{tabular}{cccc}
\hline
Methods (Accuracy/Attack success rate) & {label flipping }                     & {backdoor attack }                    & {hybrid attack }                      \\ \hline
AUROR{\cite{auror} }    & 82.21\%/43.29\%                       & 76.13\%/24.73\%                       & 80.25\%/46.81\%                       \\
FoolsGold{\cite{fung2018mitigating} }& 85.43\%/18.39\%                       & 84.49\%/19.72\%                       & 83.09\%/18.42\%                       \\
Krum{\cite{mcmahan2017communication} }     & 84.67\%/16.42\%                       & 82.65\%/{\textcolor[rgb]{1,0,0}{5.47\%}}                        & 82.16\%/11.67\%                        \\
Our method      & {\textcolor[rgb]{1,0,0} {87.43\%/4.32\%}} & {\textcolor[rgb]{1,0,0} {85.76\%}}/7.72\% & {\textcolor[rgb]{1,0,0} {84.29\%/6.87\%}} \\ \hline
\end{tabular}
\end{table*}

\begin{figure*}[t]
    \centering
    \subfloat[ ]
    {
        \includegraphics[width=3in]{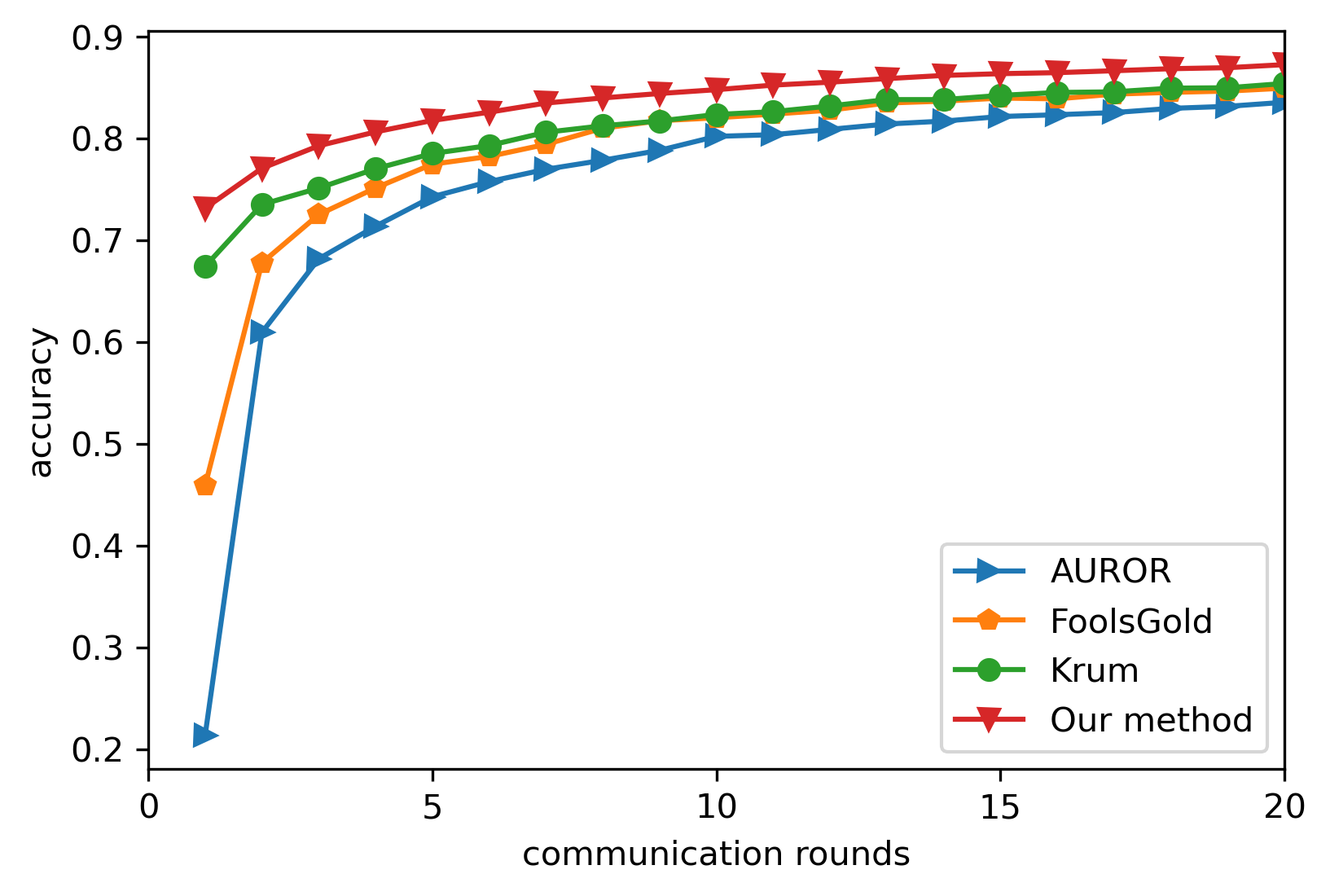}
         \label{fig4-1}
    }
    \hfil
    \subfloat[ ]
    {
	\includegraphics[width=3in]{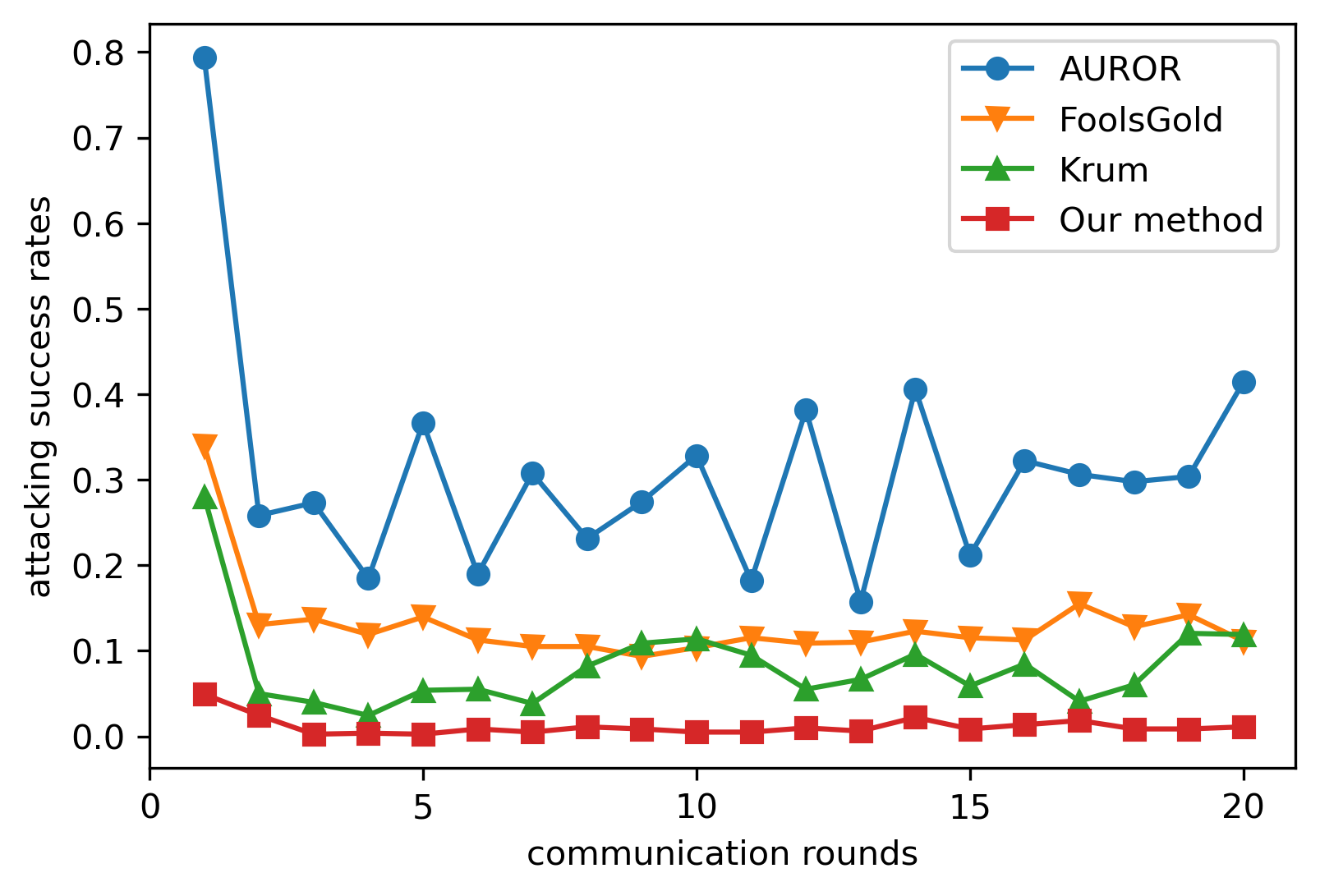}
        \label{fig4-2}
    }
    \quad    %用 \quad 来换行
    \subfloat[ ]
    {
        \includegraphics[width=3in]{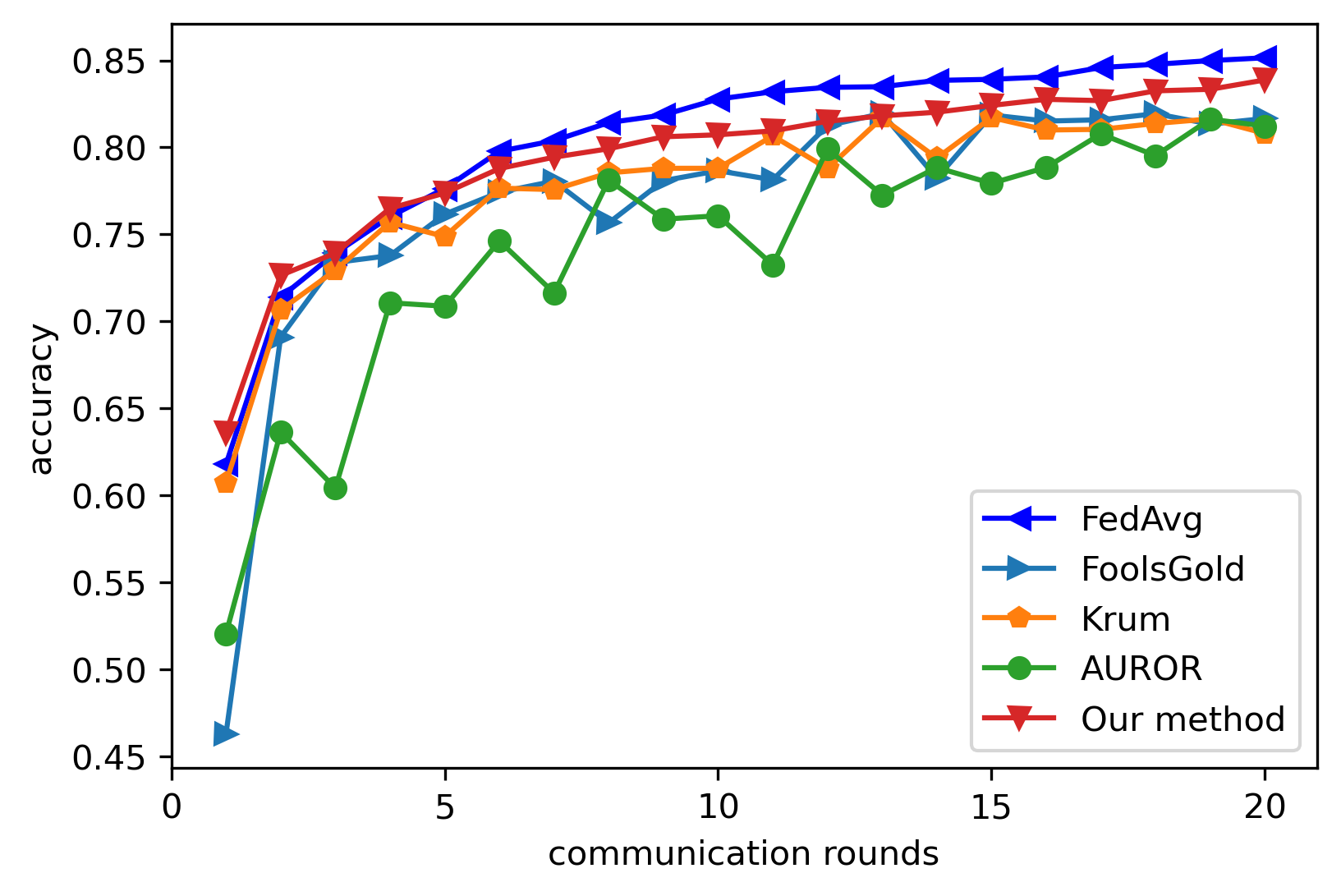}
        \label{fig4-3}
    }
    \hfil
    \subfloat[ ]
    {
	\includegraphics[width=3in]{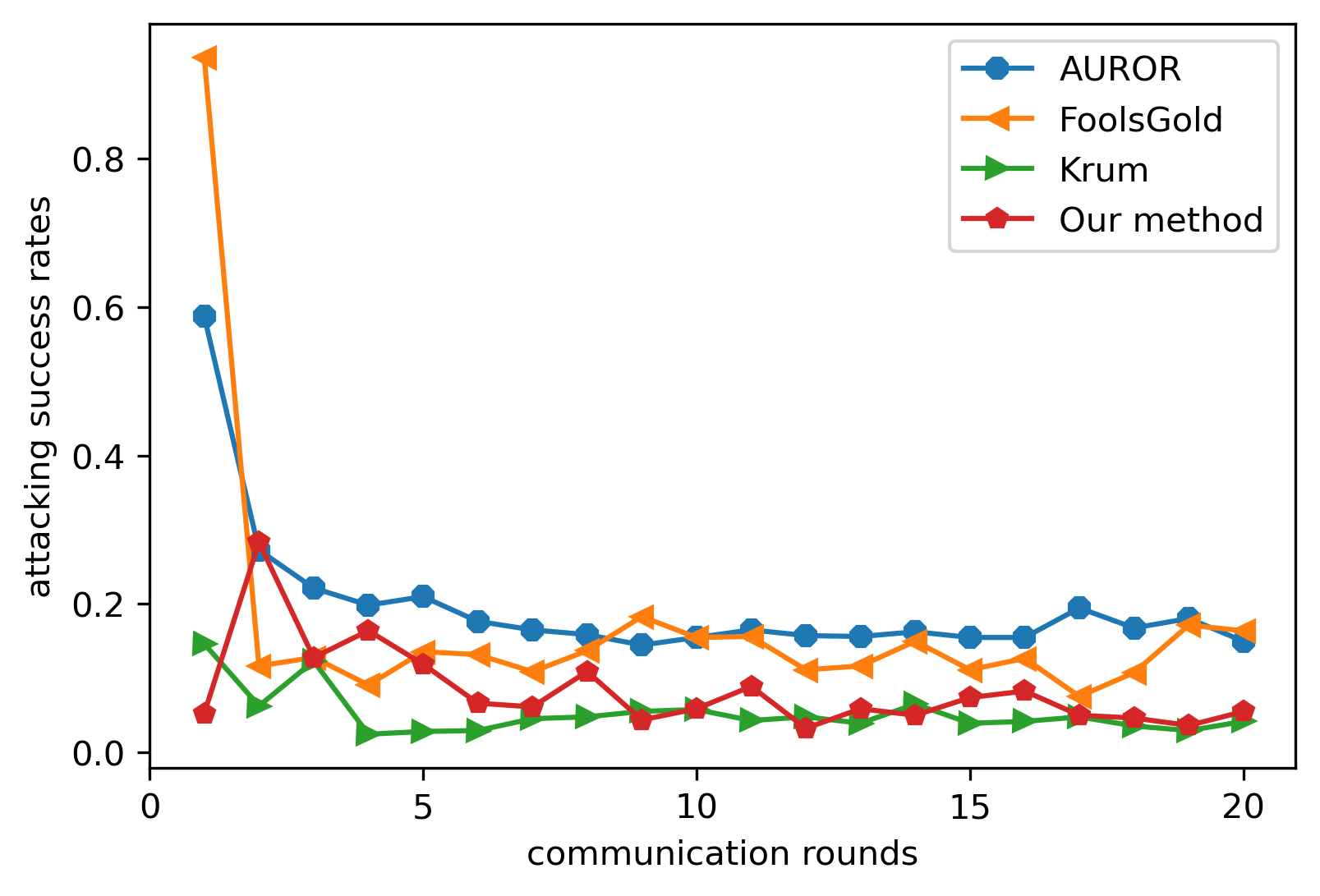}
         \label{fig4-4}
    }
    \caption{The defense methods against data poisoning attack on MNIST and ChestMNIST dataset. \ref{fig4-1} accuracy of defense methods on MNIST, \ref{fig4-1} attack success rate of defense methods on MNIST, \ref{fig4-3} accuracy of defense methods on ChestMNIST, \ref{fig4-4} attack success rate of defense methods on ChestMNIST.}
    \label{fig.4}
\end{figure*}

\subsection{Experiment and Results Analysis}

The experiments are designed for testing the proposed defense mechanism. In the label flipping attack, label ‘3’ is flipped to   
label‘2’for the MNIST, and label ‘Mass’ is flipped to label ‘Nodule’ for the ChestMNIST. As for the backdoor attacks, aim to change the small size of pixels to change the data feature contribution.

\textbf{Effectiveness.}
The first part of our evaluation assesses the effectiveness
of our methods against three kinds of data poisoning. 
For each attack, we measure the effectiveness of the model in terms of accuracy and the attack success rate.
The higher accuracy means that the main task of the attacked model is normal and efficient. And the lower attack success rate means that the attacked model is well protected by the defense.

Table {\ref{MNIST}} and Table {\ref{MedMNIST}}
present a convincing comparison between the other defense methods. Our method achieves five championships on each of the six circuits.

As we can see that the fig.{\ref{fig.4}}, our method keeps a high accuracy of all the federated learning communication rounds. 
Our method achieves the highest accuracy ($92.43\%$) among the four methods against label flipping attacks on the MNIST dataset, and our method achieves $86.23\%$ and $87.35\%$ accuracy against the backdoor attacks and hybrid attacks.
This result can explain that federated learning with the data sanitization is without compromising model accuracy by reducing the number of training samples. Fig.{\ref{fig.4}}c has shown that comparing our method to Non-protection FedAvg in the backdoor attacks, we have achieved very close results.

As for the effectiveness of defense against the attacks, our method achieves better performance than the three competitors, we respectively defend against the label flipping and obtained $2.32\%$ of the lowest attack success rate on the MNIST dataset.
Although we do not beat the Krum in backdoor attack defense($4.33\%:3.47\%$). we achieve better performance on the whole.

\textbf{Generality.}
Generality has been regarded as the biggest opponent of the model's accuracy in specificity.
To demonstrate the out-performance Generality of our method, we further do a series evaluation on the ChestMNIST, which is much harder than the MNIST dataset to perform well. We choose ResNet18 as our base network in federated learning.

Fig.{\ref{fig.4}}c proves that our method has a good Generality when meets a more complex task. On the ChestMNIST dataset, our method achieves $87.43\%$ of the highest accuracy among the four methods against label flipping attacks on the ChestMNIST dataset, and our method achieves $85.76\%$ and $84.29\%$ the accuracy against the backdoor attacks and  hybrid attacks. Yet we achieve the lowest attack success rate($4.32\%$) against the label flipping attack, $7.72\%$ against the backdoor attack, and $6.87\%$ against the hybrid attack.

Although our method filter out the poisoned data causing a reduction of training samples,
the experiments on ChestMNIST prove our method
without compromising the model.

\begin{table}[t]
\centering
\caption{Defense  against the attacks on MNIST 
 against the different ratios of adversaries to the number of participating clients.}
\label{rate}
\begin{tabular}{l|lll}
\hline
                 & 25\%  & 50\%  & 75\% \\ \hline
label attacks    & 91.23\%/3.42\% & 89.14\%/4.21\% & 87.14\%/7.21\% \\
backdoor attacks & 84.23\%/5.34\% & 82.45\%/7.42\% & 80.67\%/9.36\% \\ \hline
\end{tabular}
\end{table}

\textbf{Rate of attackers.}
To explore the impact of defense against data poisoning attacks.
We evaluate our method against $25\%$, $50\%$,
$75\%$ the different ratios of adversaries to the number of participating clients.
It is prominent that our method successfully defends against the attack by various ratios of adversaries to the number of participating clients.
As shown in Table{\ref{rate}}, with more adversaries attending to the attack, the performance of our method is somewhat degraded, but it still maintains relatively good results.

\begin{figure}[t]
\centering
\includegraphics[width=0.4 \textwidth]{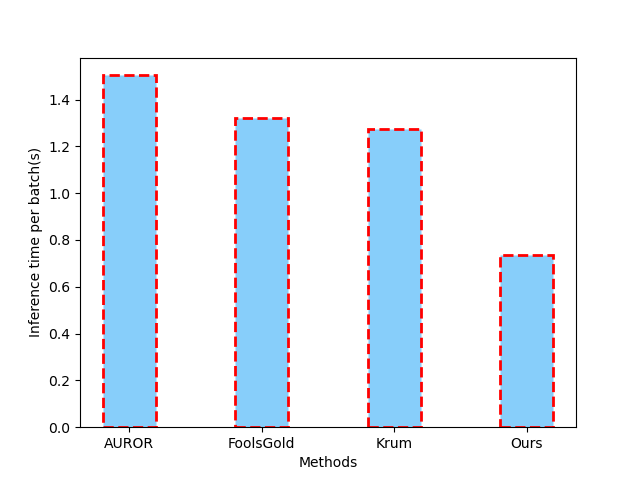}
\caption{Inference time per batch(s) of different defense methods in the training stage.}
\label{time}
\end{figure}

\textbf{Time Consumption.}
We also evaluate the time consumption among these methods.
Because our method can provide a secure data environment for federated learning, it achieves the best results on the time consumption of federated model inference time per batch.
Our method differs from existing methods in using data sanitization to protect the model from attacks.
Although our method requires extra time for data sanitization, the experimental results prove that our method defends against attacks more effectively.

\section{Conclusion}
In this paper, we introduce a Federated Data Sanitization
Defense to protect the healthcare IoMT systems from data poisoning attacks. 
Our innovative use of clustering data by features improves data security in federated learning.
We investigate the blending correlations between the poisoned data and the normal data and proposed a practical data sanitization strategy, and apply it to federated learning.
Our method does not increase the computational overhead of model training, which is especially important for large-scale data training tasks in federated learning.

\section*{Acknowledgments}
The authors would like to thank the anonymous reviewers for their thorough and constructive comments that have helped improve the quality of this article.



\bibliographystyle{IEEEtran}
\bibliography{IEEEfull}

\end{document}

%% file: math_commands.tex
\usepackage{amsmath,amsfonts,bm}

\def\eqref#1{equation~\ref{#1}}
\def\1{\bm{1}}

\def\rk{{\textnormal{k}}}

\def\mA{{\bm{A}}}

\def\mU{{\bm{U}}}

\DeclareMathAlphabet{\mathsfit}{\encodingdefault}{\sfdefault}{m}{sl}
\SetMathAlphabet{\mathsfit}{bold}{\encodingdefault}{\sfdefault}{bx}{n}

\def\gD{{\mathcal{D}}}

\def\gK{{\mathcal{K}}}

\def\gP{{\mathcal{P}}}

\def\gS{{\mathcal{S}}}

\def\gU{{\mathcal{U}}}

\DeclareMathOperator*{\argmax}{arg\,max}